\documentclass[usenatbib,usegraphicx]{mn2e}
\usepackage{amsfonts}
\usepackage{amsmath}
\usepackage{amssymb}
\usepackage{comment}
\usepackage{color}

\title[The insignificance of major mergers at high redshift]
    {The insignificance of major mergers in driving star formation at z$\simeq$2 \vspace{-0.2in}}

\author[Sugata Kaviraj et al.]
{S. Kaviraj$^{1,2}$, S. Cohen$^{3}$, R. A. Windhorst$^{3}$, J.
Silk$^{4,2}$, R. W. O'Connell$^{5}$,
\newauthor M. A. Dopita$^{6,7}$, A. Dekel$^{8}$, N. P. Hathi$^{9}$, A. Straughn$^{10}$ and M. Rutkowski$^{3}$\\\\
$^{1}$Blackett Laboratory, Imperial College London, London SW7 2AZ, UK\\
$^{2}$Department of Physics, University of Oxford, Keble Road,
Oxford, OX1 3RH, UK\\
$^{3}$School of Earth and Space Exploration, Arizona State
University, Tempe, AZ 85287-1404, USA\\
$^{4}$Institut d'Astrophysique de Paris, 98 bis boulevard Arago,
75014 Paris\\
$^{5}$Department of Astronomy, University of Virginia,
Charlottesville, VA 22904-4325, USA\\
$^{6}$Research School of Physics and Astronomy, The Australian
National University, ACT 2611, Australia\\
$^{7}$King Abdulaziz University, Astronomy Department, Faculty of Science, Jeddah, Saudi Arabia\\
$^{8}$Racah Institute of Physics, The Hebrew University, Jerusalem
91904, Israel\\
$^{9}$Carnegie Observatories, 813 Santa Babara Street, Pasadena,
California, 91101, USA\\
$^{10}$Astrophysics Science Division, Goddard Space Flight Center,
Code 665, Greenbelt, MD 20771, USA \vspace{-0.2in}}

\begin{document}

\voffset -0.5in \addtolength{\textheight}{0.5in}

\voffset -0.8in

\maketitle

\def \aj {AJ}
\def \mnras {MNRAS}
\def \pasp {PASP}
\def \apj {ApJ}
\def \apjs {ApJS}
\def \apjl {ApJL}
\def \aap {A\&A}
\def \nat {Nature}
\def \araa {ARAA}
\def \iaucirc {IAUC}
\def \aaps {A\&A Suppl.}
\def \qjras {QJRAS}
\def \na {New Astronomy}
\def\lesssim{\mathrel{\hbox{\rlap{\hbox{\lower4pt\hbox{$\sim$}}}\hbox{$<$}}}}
\def\gtrsim{\mathrel{\hbox{\rlap{\hbox{\lower4pt\hbox{$\sim$}}}\hbox{$>$}}}}


\begin{abstract}
We study the significance of major-merger-driven star formation in
the early Universe, by quantifying the contribution of this
process to the total star formation budget in 80 massive
(M$_*>10^{10}$ M$_{\odot}$) galaxies at $z \simeq 2$. Employing
visually-classified morphologies from rest-frame $V$-band HST
imaging, we find that 55$^{\pm14}$\% of the star formation budget
is hosted by \emph{non-interacting} late-types, with 27$^{\pm8}$\%
in major mergers and 18$^{\pm6}$\% in spheroids. Given that a
system undergoing a major merger continues to experience star
formation driven by other processes at this epoch (e.g. cold
accretion, minor mergers), $\sim$27\% is an \emph{upper limit} to
the major-merger contribution to star formation activity at this
epoch. The ratio of the average specific star formation rate in
major mergers to that in the non-interacting late-types is
$\sim$2.2:1, suggesting that the enhancement of star formation due
to major merging is typically modest, and that just under half the
star formation in systems experiencing major mergers is unrelated
to the merger itself. Taking this into account, we estimate that
the actual major-merger contribution to the star formation budget
may be as low as $\sim$15\%. While our study does not preclude a
major-merger-dominated era in the very early Universe, if the
major-merger contribution to star formation does not evolve
strongly into larger look-back times, then this process has a
relatively insignificant role in driving stellar mass assembly
over cosmic time.
\end{abstract}




\section{Introduction}
Understanding the processes that build massive galaxies is a
central topic in observational cosmology. The observed peak in the
cosmic star formation rate (SFR) at $z\simeq2$
\citep[e.g.][]{Madau1998,Hopkins2006} indicates that a significant
fraction of the stellar mass in today's massive galaxies is likely
to have formed around this epoch. However, the principal
mechanisms that created this stellar mass remain unclear. Was the
star formation driven by vigorous, major-merger (mass ratios $>$
1:3) induced starbursts? Or were processes other than major
mergers - e.g. cold accretion, minor mergers, etc. - responsible
for creating the bulk of the stars in today's massive galaxies, as
suggested by recent theoretical work
\citep[e.g.][]{Keres2009,Dekel2009a}?

Modern surveys that access large UV/optically-selected samples of
galaxies at $z>1.5$ have facilitated the empirical study of star
formation around $z \simeq 2$
\citep[e.g.][]{Daddi2004,Erb2006,Reddy2005,Daddi2007,Santini2009,Hathi2010,Wuyts2011}.
Star-forming galaxies at this epoch lie on a star-formation `main
sequence' \citep[e.g.][]{Daddi2007,Reddy2012}, where galaxy SFRs
are proportional to their stellar masses with a slope of unity
(relatively passive galaxies lie below this sequence). The growing
body of observational work on these galaxies increasingly suggests
that much of the cosmic star formation at this epoch may be
unrelated to the major-merger process. Integral-field spectroscopy
of star-forming galaxies around $z \simeq2$ has revealed a high
fraction of systems with properties indicative of turbulent disks
and only a modest incidence of major mergers \citep[e.g.][see also
Law et al. 2009, van Dokkum et al.
2011]{ForsterSchreiber2006,Genzel2008,Shapiro2008,ForsterSchreiber2009,Cresci2009,Genzel2011,Mancini2011}.
Imaging studies, that have explored the rest-frame UV and optical
morphologies of star-forming galaxies at these epochs
\cite[e.g.][]{Lotz2006,ForsterSchreiber2011,Law2012}, have also
indicated a preponderance of non-merging systems amongst
high-redshift star formers, suggesting that the role of major
mergers may indeed be subordinate to that of other processes in
driving star formation at this epoch.

\begin{figure}
\begin{center}
$\begin{array}{c}
\includegraphics[width=2.2in]{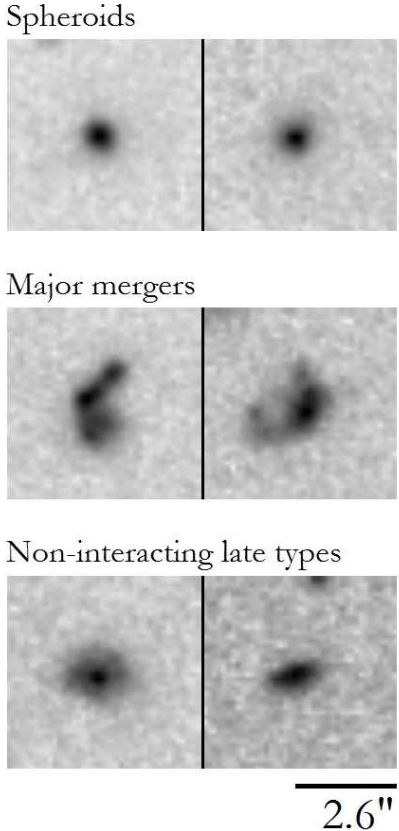}
\end{array}$
\caption{Example J+H composite images for the morphological
classes into which we split our galaxy sample. Galaxies are
classified into spheroids (top row), major mergers (disturbed
systems with multiple nuclei and clear, extended tidal features,
middle row) and non-interacting late-types (bottom row).
{\color{black}An angular scale which spans half the image is
shown.} The images are $\sim$45 kpc on a side. Note that the
images usually appear better on screen than in print.
\vspace{-0.1in}} \label{fig:sfr_mass}
\end{center}
\end{figure}

In a recent study, \citet{Rodighiero2011} have shown that
`starbursts' -- systems that show enhanced star formation and lie
off the main sequence of normal star-forming galaxies -- have a
relatively minor role at this epoch, accounting for around 10\% of
the cosmic star formation activity. However, starbursts can be
driven either via major mergers or by dense nuclear star-forming
regions \citep[e.g.][]{DiMatteo2007,Dekel2009a,Daddi2010}. More
importantly, many major mergers share the same star-formation
characteristics as normal star-forming galaxies at this epoch
\cite[e.g.][see also Di Matteo et al. 2007, Kaviraj et al.
2012]{Law2012} and thus lie on the main star-formation sequence
itself. As a result, a unique one-to-one mapping is unlikely to
exist between major mergers and starbursts. To probe the relative
significance of major-merger-driven star formation at $z \simeq2$,
it is desirable to quantify the proportion of the total star
formation budget that is attributable to systems that are
\emph{morphologically selected} as major mergers at this epoch.
This has not been directly addressed by previous work and
represents both a quantitative empirical result and a useful
constraint on theoretical models at high redshift.

Deep near-infrared imaging from current WFC3 surveys -- which
trace rest-frame optical wavelengths at $z \simeq2$ -- enables us
to morphologically classify massive galaxies at this epoch and
study how star formation activity is apportioned in terms of
galaxy morphology (e.g. major mergers, non-interacting late-types,
etc). It is worth noting, however, that a system undergoing a
major merger at $z \simeq2$ continues to experience star formation
driven by gas inflow via other processes such as cold accretion
and minor mergers (major mergers can be thought of as simply the
`clumpiest' part of the material flowing in along the cosmic web).
Both theoretical \citep[e.g.][]{Dekel2009a} and observational
\citep[e.g.][]{Kaviraj2012,Law2012} work indicates that star
formation due to these other processes is significant at this
epoch and possibly comparable to the major-merger-driven activity.
Hence, in addition to splitting the star formation budget by
morphology, it is necessary to consider the proportion of star
formation in major-merging systems that is unrelated to the
major-merger process, and subtract this from the fraction of the
star formation budget hosted by systems with major-merger
morphology\footnote{Note that the situation is significantly
different at low redshift, where gas-rich major mergers can
enhance star formation by orders of magnitude
\citep[e.g.][]{Mihos1996}, because secular processes drive star
formation weakly. Almost all the star formation in
\emph{low-redshift} major mergers is, therefore, attributable to
the merger itself.}.

Here, we probe these questions using a complete, rest-frame
optically-selected sample of massive (M$_*>10^{10}$ M$_{\odot}$)
galaxies at $z \simeq2$, drawn from the WFC3 Early Release Science
(ERS) programme, which provides unprecedentedly deep near-infrared
HST imaging and ten-filter photometry in the GOODS-South field.
Section 2 describes the galaxy sample that underpins this study.
In Section 3, we describe the derivation of galaxy properties e.g.
SFRs, stellar masses and internal extinctions. We study the
proportional contribution of major mergers to the total star
formation budget in Section 4 and summarise our findings in
Section 5. Throughout, we use the WMAP7 cosmological parameters
\citep{Komatsu2011} and present photometry in the AB magnitude
system \cite{Oke1983}. \vspace{-0.15in}


\vspace{-0.1in}
\section{Galaxy sample and morphological classifications}
The WFC3 ERS programme has imaged $\sim$45 arcmin$^{2}$ of the
GOODS-South field in the WFC3 UVIS (F225W, F275W, F336W) and IR
(F098M [Y], F125W [J], F160W [H]) channels, with exposure times of
1-2 orbits per filter. The observations, data reduction, and
instrument performance are described in detail in
\citet{Windhorst2011}. Together with the existing ACS \emph{BViz}
imaging \citep{Giavalisco2004}, the data provide 10-band
panchromatic coverage over 0.2 - 1.7 $\mu$m, with 5$\sigma$ point
source depths of $AB\lesssim 26.1-26.4$ mag in the UV and
$AB\lesssim27.2-27.5$ mag in the IR.

Here, we focus on an $H$-band selected sample of 80 ERS galaxies,
that have stellar masses M$_*>10^{10}$M$_{\odot}$ and redshifts in
the range $1.9<z<2.1$. {\color{black}For 12\% if our sample (10
galaxies) we use published spectroscopic redshifts from
\citet{Santini2009,Popesso2009,Ferreras2009}. For the remaining
objects we use photometric redshifts, calculated using the EAZY
code (Brammer et al. 2008). The peak of the redshift probability
distribution from EAZY is used as the best estimate of the
redshift - the accuracy of EAZY redshifts at this epoch is $\Delta
z \sim 0.1$ and the nominal time interval our defined by redshift
range is $\sim$0.3 Gyr.} The $H$-band traces rest-frame $V$ at $z
\simeq2$ and the galaxy sample is complete within these stellar
mass and redshift ranges \citep{Windhorst2011}. Studying the
massive end of the galaxy population restricts us to systems that
are both bright ($H (AB) < 24.2$ mag) and extended, which
facilitates reliable morphological classification.The narrow
redshift interval minimises both morphological K-corrections and
overlap between the spheroid and major-merger morphological
classes (as we discuss in Section 4).

Here we classify galaxies via visual inspection of their composite
J+H images. Since the J and H filters correspond to the rest-frame
optical wavelengths at $z \simeq 2$, these images trace the
underlying stellar populations in each galaxy and not just the
UV-emitting star-forming regions. Visual classification of
morphologies in the high-redshift Universe has been commonly
employed in the literature, using rest-frame optical HST images
that have similar or fainter surface-brightness limits compared to
the ERS images used here
\citep[e.g.][]{Windhorst2002,Cassata2010,Kaviraj2011,Cameron2011,Kocevski2012,Law2012}.
Visual classification offers better precision and consistency than
morphological parameters \citep[such as CAS, M$_{20}$, Gini
coefficient, see
e.g.][]{Abraham1996,Conselice2003,Lotz2004,Taylor-Mager2007},
which can be more sensitive to image resolution and
signal-to-noise \citep[e.g.][]{Lisker2008,Kartaltepe2010} but are
valuable for classifying large datasets, where visual
classification is prohibitively time-consuming.

Galaxies are classified into the following three broad
morphological classes: [1] spheroids [2] non-interacting
late-types and [3] major mergers, which are disturbed systems that
exhibit multiple nuclei and clear, \emph{extended} tidal features.
The number fractions in classes [1], [2] and [3] are 19\%, 43\%
and 38\% respectively. Figure 1 presents examples of objects drawn
from each morphological class.


\vspace{-0.2in}
\section{Parameter estimation: stellar masses and intrinsic star formation rates}
A variety of methods have been employed in the literature to
derive galaxy SFRs. Calibrations (often based on samples of local
galaxies) can be used to convert X-ray, UV, infrared or radio
luminosities into estimates of SFR \citep[see
e.g.][]{Reddy2004,Daddi2007,Pannella2009,Elbaz2011}.
Alternatively, galaxy spectral energy distributions (SEDs) can be
fitted to theoretical star formation histories (SFHs) to derive
SFRs \citep[e.g.][]{Shapley2005,Law2012}. Since dust is included
as a free parameter in these SFHs, intrinsic (i.e. dust-corrected)
SFRs can be derived self-consistently using this method. The
derivation of reliable SFRs ideally requires rest-frame UV
photometry (as is the case here), since the leverage in the SFR
and the dust extinction comes largely from these wavelengths.

In this paper, we calculate galaxy SFRs via SED fitting. The
WFC3/ACS photometry of each individual galaxy is compared to a
large library of synthetic photometry, constructed using constant
SFHs, each described by a stellar mass ($M$), age ($T$),
metallicity ($Z$) and internal extinction ($E_{B-V}$). We vary $T$
between 0.05 Gyrs and the look-back time to $z=20$ in the
rest-frame of the galaxy, $Z$ between 0.1 Z$_{\odot}$ and 2.5
Z$_{\odot}$ and $E_{B-V}$ between 0 and 1 mag. Synthetic
magnitudes are generated by folding the model SFHs with the
stellar models of \citet{BC2003}, with dust attenuation applied
following \citet{Calzetti2000}. The likelihood of each model,
$\exp (-\chi^2/2)$, is calculated using the value of $\chi^2$,
computed in the standard way. Estimates for parameters such as
stellar mass, internal extinction and SFR are derived by
marginalising each parameter from the joint probability
distribution, to extract its one-dimensional probability density
function (PDF). We use the median of this PDF as the best estimate
of the parameter in question, with the 25 and 75 percentile values
(which enclose 50\% of the probability) yielding an associated
uncertainty. Since dust is explicitly taken into account in this
process, we derive \emph{intrinsic} SFRs, free of internal
reddening, directly from the SED fitting process. The derived
internal extinctions, SFRs and stellar masses are uncertain by
$\sim$0.1 mag, $\sim$0.1 dex and $\sim$0.2 dex respectively.

\begin{figure}
\begin{center}
$\begin{array}{c}
\includegraphics[width=3in]{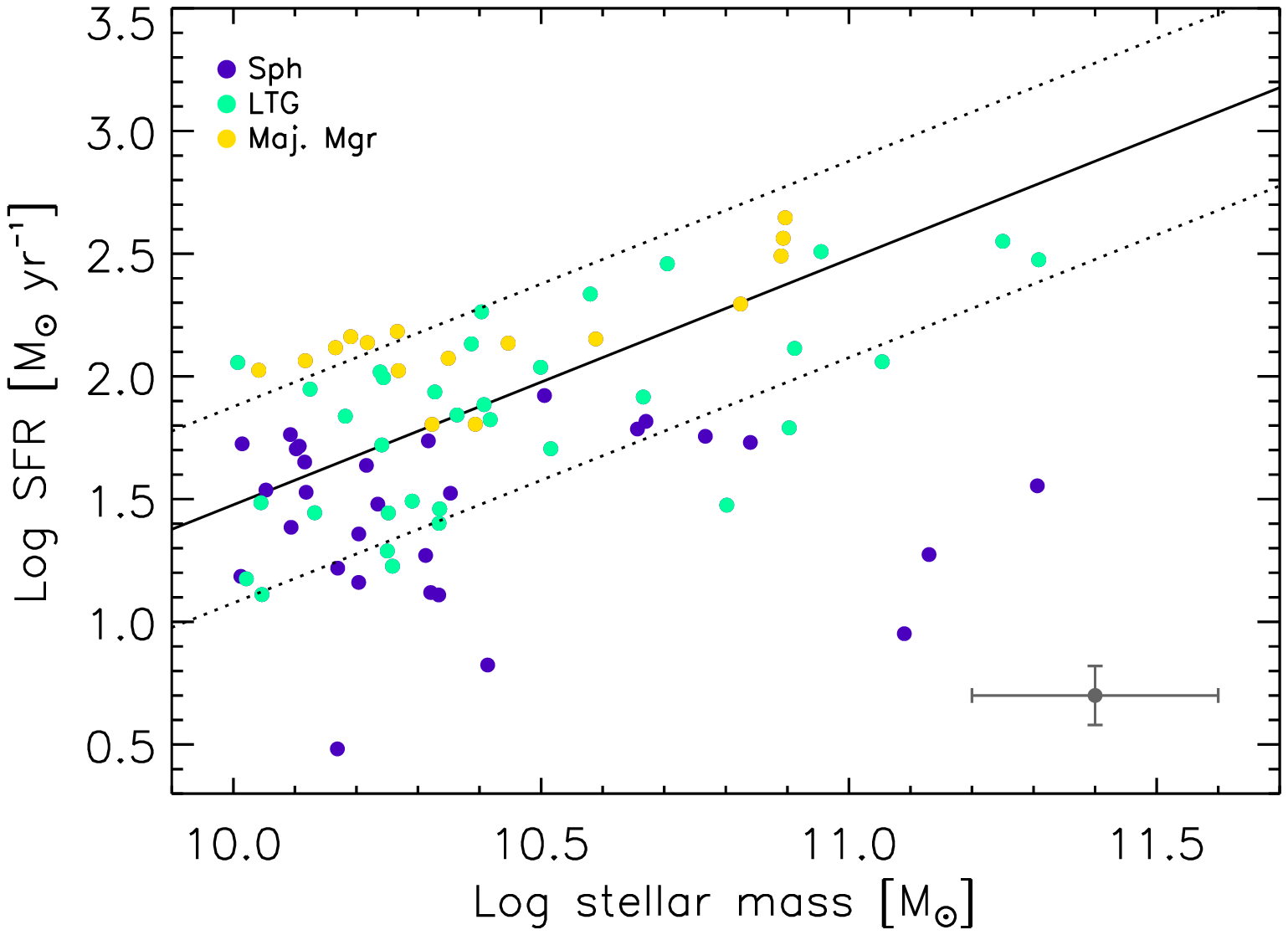}\\
\includegraphics[width=3in]{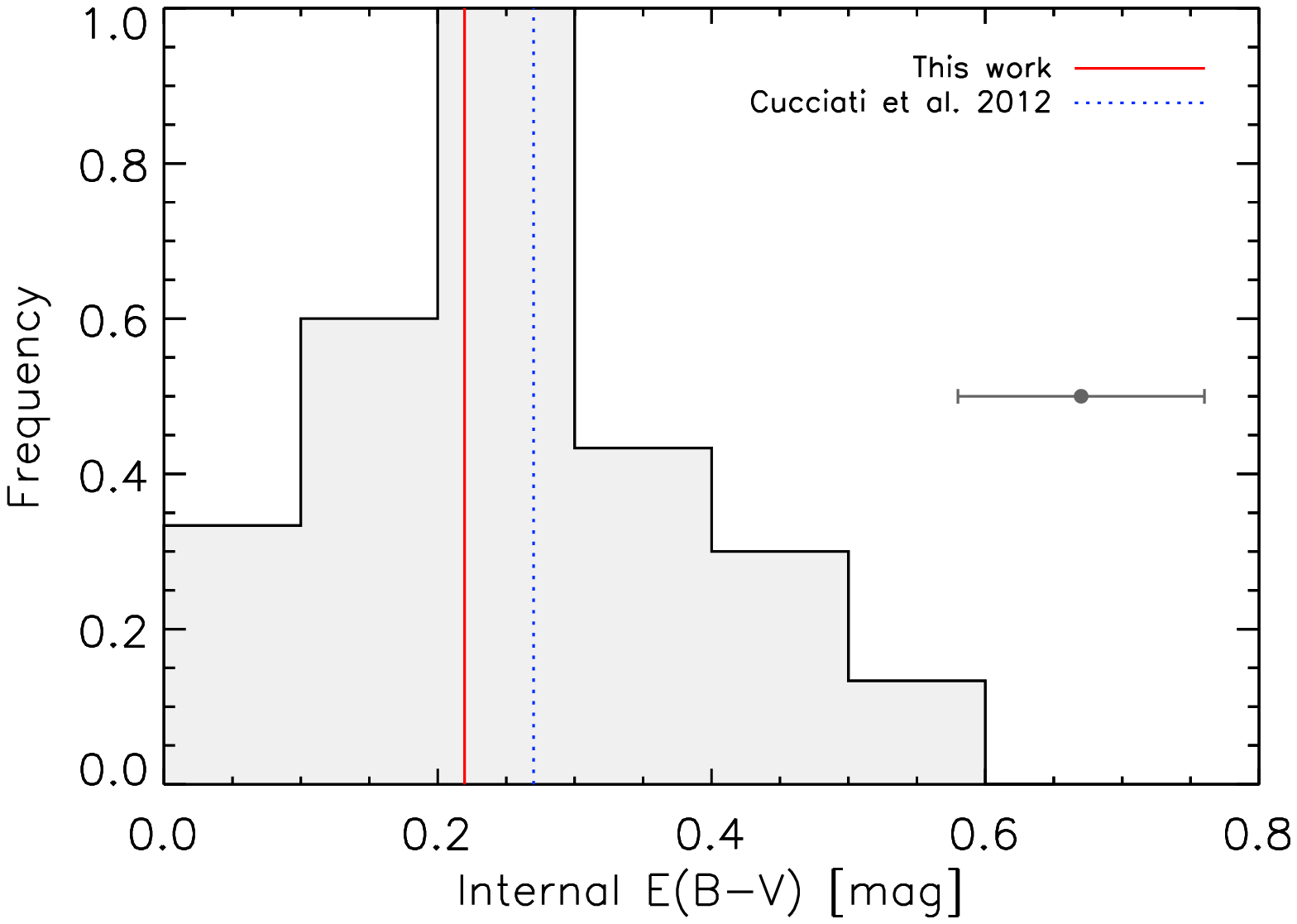}
\end{array}$
\caption{TOP: SFR as a function of galaxy stellar mass in our
galaxy sample. The main sequence of star-forming galaxies at $z
\simeq2$ is shown using the solid line (the observed scatter is
indicated using the dotted lines). Galaxy morphologies are shown
colour-coded (black = spheroids, blue = non-interacting
late-types, red = major mergers). BOTTOM: The distribution of
derived extinction values for our galaxies. The median value is
shown by the vertical red line and the median from the recent
literature (see Tresse et al. 2007 and Cucciati et al. 2012) is
shown by the blue dotted line.} \label{fig:sfr_mass}
\end{center}
\end{figure}

In the top panel of Figure \ref{fig:sfr_mass}, we plot our derived
SFRs vs. galaxy stellar mass. The SFR values for our star-forming
galaxies are consistent with the star formation main sequence at
these epochs defined by the recent literature. Galaxies that lie
below this sequence are typically spheroids, which are relatively
passively-evolving systems. Recall that, unlike studies that
specifically target star-forming systems, the mass-complete sample
employed here is not biased against galaxies with low star
formation rates. In the bottom panel of Figure \ref{fig:sfr_mass},
we present the distribution of derived internal $E_{B-V}$ values
for our galaxies. The spread in our values ($0<E_{B-V} <0.5$ mag)
agrees well with that found by other studies
\citep[e.g.][]{Law2012b} and the median of our distribution
($E_{B-V}\sim0.25$ mag i.e. $A_{FUV} \sim 2.2$ assuming Calzetti
et al. 2000) is in good agreement with the literature at $z
\simeq2 $ (see e.g. \citet{Tresse2007}, \citet{Law2012b} and
\citet[][see their Figure 4]{Cucciati2012}).


\vspace{-0.15in}
\section{The major-merger contribution to the star formation budget}
We begin by exploring how star formation activity is apportioned
in terms of galaxy morphology, by summing the derived SFRs of
galaxies in each morphological class and considering the
fractional contribution of these classes to the total star
formation budget (Figure 3). We find that 55$^{\pm14}$\% of the
star formation activity takes place in non-interacting late-types,
with 27$^{\pm8}$\% in major mergers and the rest (18$^{\pm6}$\%)
in systems that have spheroidal morphology. It is worth noting
that the proportion of star formation driven by
\emph{morphologically selected} major mergers calculated here is
higher than the corresponding value derived for starbursts by
\citet{Rodighiero2011}. As we noted in the introduction, this is
due to the fact that many major mergers exhibit similar or only
modestly-enhanced SFRs compared to normal star-forming galaxies,
lie on or close to the star-forming main sequence (see Figure
\ref{fig:sfr_mass} above) and are, therefore, not part of the more
extreme starburst population.

The predominance of non-interacting late-types in the total star
formation budget indicates that major mergers are not the dominant
mechanism driving star formation in massive galaxies at $z\simeq
2$. Furthermore, as we noted in the introduction, systems
undergoing major mergers continue to experience star formation via
other processes (e.g. cold flows and minor mergers). Hence, 27\%
represents an \emph{upper limit} to the major-merger contribution
to the star formation budget. To improve our estimate, we consider
the \emph{enhancement} of star formation due to major merging,
since this better represents the portion of the star formation
activity that is directly attributable to this process. While
measuring this enhancement is not possible in individual major
mergers, we can estimate a \emph{typical} value for the population
as whole by comparing the mean specific SFR in the major mergers
to that in the non-interacting late-types.

The ratio of the mean specific SFRs in these two morphological
classes is $\sim$2.2:1 (major mergers : non-interacting
late-types), implying that, \emph{on average}, around half the
star formation in major-mergers is likely driven by other
processes. This relatively modest enhancement in star formation
activity due to major merging is consistent with the findings of
recent theoretical work \citep[e.g.][]{Cen2011} and also empirical
studies that do not find significant differences between the SFRs
of galaxies that are morphologically disturbed and those that are
not at this epoch \citep[e.g.][]{Kaviraj2012,Law2012}. Thus, if
around half the star formation in major mergers is unrelated to
the merger itself, then the major-merger contribution to the total
star formation budget is likely to be as low as $\sim$15\% (i.e.
27\% $\times$ 1.2/2.2).

Before we conclude this section, we briefly discuss the spheroid
population in the context of the major mergers. We note first that
the time interval spanned by our study ($\sim$0.3 Gyr) is shorter
than the effective timescales (0.5-2 Gyr) over which major mergers
coalesce \citep[see e.g.][]{Lotz2008,Newman2012}, so that the
morphological classes do not overlap with each other. More
importantly, however, \citet{Kaviraj2012} have used
high-resolution cosmological simulations to demonstrate that
spheroids at $1<z<3$ that are remnants of \emph{recent} major
mergers (i.e. ones that coalesced within the last $\sim$0.5 Gyr)
will exhibit clear tidal features at the depth of the ERS images.
This study has further demonstrated that many newborn spheroids in
this redshift range do \emph{not} carry such morphological
disturbances, indicating that a significant fraction of these
systems are not built via major mergers (in agreement with the
results of recent theoretical work, e.g. Dekel et al. 2009).
Around 15\% of the spheroids in our sample show morphological
disturbances and these galaxies account for $\sim$3\% of the
\emph{total} star formation budget. While the spheroid and major
merger classes do not overlap (as discussed above), it is clear
that, even if we added the disturbed spheroids to the major merger
portion of the star formation budget, our conclusions would remain
unchanged. Our analysis therefore indicates that major mergers
contribute a relatively insignificant fraction ($\sim$15\%) of the
total star formation budget in massive galaxies at $z \simeq2$ and
are not the principal driver of cosmic star formation at this
epoch. \vspace{-0.1in}

\begin{figure}
  \begin{minipage}{0.5\textwidth}
    \begin{center}
        \begin{tabular}{l|c}

            Morphology & Fraction of SF budget\\ \hline
            Non-interacting late-types & 0.55$^{\pm 0.14}$\\
            Major merger               & 0.27$^{\pm 0.08}$\\
            Spheroids                  & 0.18$^{\pm 0.06}$

        \end{tabular}
    \end{center}
  \end{minipage}
  \begin{minipage}{0.5\textwidth}
     \begin{center}
    \includegraphics[width=2.5in]{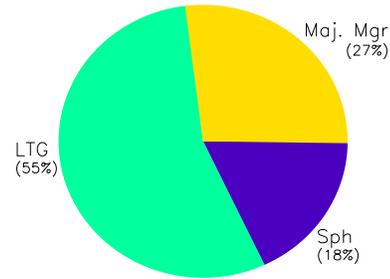}
    \end{center}
  \caption{TOP: The fraction of the total star formation budget in
massive (M$_*>10^{10}$ M$_{\odot}$) systems at $z \simeq2$ that is
hosted by various morphological types. Major mergers account for
less than a third (27\%) of the total star formation budget, while
non-interacting late-type galaxies host more than half of the star
formation activity. Given that systems undergoing major mergers
continue to experience star formation driven by other processes,
27\% is an \emph{upper limit} to the major merger contribution to
the star formation budget. The actual contribution is likely to be
$\sim$15\% of the total star formation budget (see text in Section
4 for details). BOTTOM: A pie-chart visualisation of the star
formation budget apportioned in terms of galaxy morphology (Sph =
Spheroids, LTG = Non-interacting late-type galaxies, Maj. Mgr =
Major mergers).}
\end{minipage}
\label{fig:sf_budget}
\end{figure}


\vspace{-0.15in}
\section{Summary}
We have explored the significance of major mergers in driving star
formation at high redshift, by quantifying the contribution of
this process to the total star formation budget in a sample of 80
massive (M$_*>10^{10}$ M$_{\odot}$) galaxies at $z \simeq 2$. We
have found that 55$^{\pm14}$\% of the total star formation
activity in massive galaxies at this epoch is hosted by late-type
galaxies that are not interacting with other systems, with
27$^{\pm8}$\% in major mergers, and the rest in spheroids.

Since systems undergoing major mergers continue to experience star
formation driven by other processes (e.g. cold flows and minor
mergers), $\sim$27\% is an \emph{upper limit} to the contribution
of major mergers to the total star formation budget. To improve
our estimate, we have considered the typical enhancement of star
formation induced by a major merger, since this is the portion of
star formation activity which is directly attributable to this
process. We have estimated this enhancement using the ratio of the
mean specific star formation rate in the major merger population
to that in the non-interacting late-type galaxies. In agreement
with recent observational and theoretical work, we have found a
relatively modest enhancement ($\times$2.2), which implies that,
on average, just under half the star formation activity in major
mergers is unrelated to the merger itself. This reduces the
contribution of major mergers to the total star formation budget
in massive galaxies at this epoch to $\sim$15\%. Our analysis
therefore indicates that the contribution of major mergers to the
total star formation budget in massive galaxies at $z \simeq2$ is
relatively insignificant and this process is not the principal
driver of cosmic star formation at this epoch.

Since our study is based on instantaneous star formation rates, it
provides only a snapshot of the star formation budget in massive
galaxies at $z \simeq2$. Thus, while our data cannot rule out a
merger-dominated era in the very early Universe, if the
major-merger contribution to stellar mass assembly does not evolve
strongly into earlier look-back times ($z>2$), then major mergers
are unlikely to be significant contributors to the overall buildup
of stellar mass in the Universe. In forthcoming papers we will use
morphological analyses of large datasets such as CANDELS -- e.g.
via projects such as Galaxy Zoo \citep{Lintott2008}, which uses
450,000+ members of the public to visually classify large survey
datasets -- to comprehensively study the contribution of major
mergers to the star formation budget as a function of stellar
mass, environment and redshift.


\vspace{-0.2in}
\section*{Acknowledgements}
{\color{black}We are grateful to the anonymous referee for many
constructive comments.} Richard Ellis, Emanuele Daddi, Naveen
Reddy, Giulia Rodighiero and David Law are thanked for discussions
and constructive comments. SK acknowledges fellowships from the
1851 Royal Commission, Imperial College and Worcester College
Oxford. We are grateful to the Director of STScI for awarding
Director's Discretionary time for the WFC3 ERS programme. RWO
acknowledges partial support from NASA grant GO-11359. RAW
acknowledges NASA JWST Interdisciplinary Scientist grant
NAG5-12460. AD acknowledges ISF grant 6/08, GIF grant
G-1052-104.7/2009, DIP grant STE1869/1-1.GE625/15-1 and NSF grant
AST-1010033.

\nocite{Brammer2008,Law2009,VD2011}


\vspace{-0.2in}
\small
\bibliographystyle{mn2e}
\bibliography{references}

\begin{thebibliography}{}

\bibitem[\protect\citeauthoryear{{Abraham}, {Tanvir}, {Santiago}, {Ellis},
  {Glazebrook} \& {van den Bergh}}{{Abraham} et~al.}{1996}]{Abraham1996}
{Abraham} R.~G.,  {Tanvir} N.~R.,  {Santiago} B.~X.,  {Ellis} R.~S.,
  {Glazebrook} K.,    {van den Bergh} S.,  1996, MNRAS, 279, L47

\bibitem[\protect\citeauthoryear{{Brammer}, {van Dokkum} \& {Coppi}}{{Brammer}
  et~al.}{2008}]{Brammer2008}
{Brammer} G.~B.,  {van Dokkum} P.~G.,    {Coppi} P.,  2008, \apj, 686, 1503

\bibitem[\protect\citeauthoryear{{Bruzual} \& {Charlot}}{{Bruzual} \&
  {Charlot}}{2003}]{BC2003}
{Bruzual} G.,  {Charlot} S.,  2003, MNRAS, 344, 1000

\bibitem[\protect\citeauthoryear{{Calzetti}, {Armus}, {Bohlin}, {Kinney},
  {Koornneef} \& {Storchi-Bergmann}}{{Calzetti} et~al.}{2000}]{Calzetti2000}
{Calzetti} D.,  {Armus} L.,  {Bohlin} R.~C.,  {Kinney} A.~L.,  {Koornneef} J.,
    {Storchi-Bergmann} T.,  2000, ApJ, 533, 682

\bibitem[\protect\citeauthoryear{{Cameron}, {Carollo}, {Oesch}, {Bouwens},
  {Illingworth}, {Trenti}, {Labb{\'e}} \& {Magee}}{{Cameron}
  et~al.}{2011}]{Cameron2011}
{Cameron} E.,  {Carollo} C.~M.,  {Oesch} P.~A.,  {Bouwens} R.~J.,
  {Illingworth} G.~D.,  {Trenti} M.,  {Labb{\'e}} I.,    {Magee} D.,  2011,
  \apj, 743, 146

\bibitem[\protect\citeauthoryear{{Cassata}, {} \& {et al.}}{{Cassata}
  et~al.}{2010}]{Cassata2010}
{Cassata} P.,  {}   {et al.} 2010, \apjl, 714, L79

\bibitem[\protect\citeauthoryear{{Cen}}{{Cen}}{2011}]{Cen2011}
{Cen} R.,  2011, ArXiv e-prints

\bibitem[\protect\citeauthoryear{{Conselice}, {Bershady}, {Dickinson} \&
  {Papovich}}{{Conselice} et~al.}{2003}]{Conselice2003}
{Conselice} C.~J.,  {Bershady} M.~A.,  {Dickinson} M.,    {Papovich} C.,  2003,
  AJ, 126, 1183

\bibitem[\protect\citeauthoryear{{Cresci}, {} \& {et al.}}{{Cresci}
  et~al.}{2009}]{Cresci2009}
{Cresci} G.,  {}   {et al.} 2009, \apj, 697, 115

\bibitem[\protect\citeauthoryear{{Cucciati}, {} \& {et al.}}{{Cucciati}
  et~al.}{2012}]{Cucciati2012}
{Cucciati} O.,  {}   {et al.} 2012, \aap, 539, A31

\bibitem[\protect\citeauthoryear{{Daddi}, {} \& {et al.}}{{Daddi}
  et~al.}{2004}]{Daddi2004}
{Daddi} E.,  {}   {et al.} 2004, \apjl, 600, L127

\bibitem[\protect\citeauthoryear{{Daddi}, {} \& {et al.}}{{Daddi}
  et~al.}{2007}]{Daddi2007}
{Daddi} E.,  {}   {et al.} 2007, \apj, 670, 156

\bibitem[\protect\citeauthoryear{{Daddi}, {} \& {et al.}}{{Daddi}
  et~al.}{2010}]{Daddi2010}
{Daddi} E.,  {}   {et al.} 2010, \apjl, 714, L118

\bibitem[\protect\citeauthoryear{{Dekel}, {Birnboim}, {Engel}, {Freundlich},
  {Goerdt}, {Mumcuoglu}, {Neistein}, {Pichon}, {Teyssier} \& {Zinger}}{{Dekel}
  et~al.}{2009}]{Dekel2009a}
{Dekel} A.,  {Birnboim} Y.,  {Engel} G.,  {Freundlich} J.,  {Goerdt} T.,
  {Mumcuoglu} M.,  {Neistein} E.,  {Pichon} C.,  {Teyssier} R.,    {Zinger} E.,
   2009, \nat, 457, 451

\bibitem[\protect\citeauthoryear{{Di Matteo}, {Combes}, {Melchior} \&
  {Semelin}}{{Di Matteo} et~al.}{2007}]{DiMatteo2007}
{Di Matteo} P.,  {Combes} F.,  {Melchior} A.,    {Semelin} B.,  2007, A\&A,
  468, 61

\bibitem[\protect\citeauthoryear{{Elbaz}, {} \& {et al.}}{{Elbaz}
  et~al.}{2011}]{Elbaz2011}
{Elbaz} D.,  {}   {et al.} 2011, \aap, 533, A119

\bibitem[\protect\citeauthoryear{{Erb}, {Steidel}, {Shapley}, {Pettini},
  {Reddy} \& {Adelberger}}{{Erb} et~al.}{2006}]{Erb2006}
{Erb} D.~K.,  {Steidel} C.~C.,  {Shapley} A.~E.,  {Pettini} M.,  {Reddy} N.~A.,
     {Adelberger} K.~L.,  2006, \apj, 647, 128

\bibitem[\protect\citeauthoryear{{Ferreras}, {Lisker}, {Pasquali}, {Khochfar}
  \& {Kaviraj}}{{Ferreras} et~al.}{2009}]{Ferreras2009}
{Ferreras} I.,  {Lisker} T.,  {Pasquali} A.,  {Khochfar} S.,    {Kaviraj} S.,
  2009, MNRAS, 396, 1573

\bibitem[\protect\citeauthoryear{{F{\"o}rster Schreiber}, {} \& {et
  al.}}{{F{\"o}rster Schreiber} et~al.}{2006}]{ForsterSchreiber2006}
{F{\"o}rster Schreiber} N.~M.,  {}   {et al.} 2006, \apj, 645, 1062

\bibitem[\protect\citeauthoryear{{F{\"o}rster Schreiber}, {} \& {et
  al.}}{{F{\"o}rster Schreiber} et~al.}{2009}]{ForsterSchreiber2009}
{F{\"o}rster Schreiber} N.~M.,  {}   {et al.} 2009, \apj, 706, 1364

\bibitem[\protect\citeauthoryear{{F{\"o}rster Schreiber}, {Shapley}, {Erb},
  {Genzel}, {Steidel}, {Bouch{\'e}}, {Cresci} \& {Davies}}{{F{\"o}rster
  Schreiber} et~al.}{2011}]{ForsterSchreiber2011}
{F{\"o}rster Schreiber} N.~M.,  {Shapley} A.~E.,  {Erb} D.~K.,  {Genzel} R.,
  {Steidel} C.~C.,  {Bouch{\'e}} N.,  {Cresci} G.,    {Davies} R.,  2011, \apj,
  731, 65

\bibitem[\protect\citeauthoryear{{Genzel}, {} \& {et al.}}{{Genzel}
  et~al.}{2008}]{Genzel2008}
{Genzel} R.,  {}   {et al.} 2008, \apj, 687, 59

\bibitem[\protect\citeauthoryear{{Genzel}, {} \& {et al.}}{{Genzel}
  et~al.}{2011}]{Genzel2011}
{Genzel} R.,  {}   {et al.} 2011, \apj, 733, 101

\bibitem[\protect\citeauthoryear{{Giavalisco}, {} \& {et al.}}{{Giavalisco}
  et~al.}{2004}]{Giavalisco2004}
{Giavalisco} M.,  {}   {et al.} 2004, ApJ, 600, L93

\bibitem[\protect\citeauthoryear{{Hathi}, {} \& {et al.}}{{Hathi}
  et~al.}{2010}]{Hathi2010}
{Hathi} N.~P.,  {}   {et al.} 2010, \apj, 720, 1708

\bibitem[\protect\citeauthoryear{{Hopkins} \& {Beacom}}{{Hopkins} \&
  {Beacom}}{2006}]{Hopkins2006}
{Hopkins} A.~M.,  {Beacom} J.~F.,  2006, \apj, 651, 142

\bibitem[\protect\citeauthoryear{{Kartaltepe}, {} \& {et al.}}{{Kartaltepe}
  et~al.}{2010}]{Kartaltepe2010}
{Kartaltepe} J.~S.,  {}   {et al.} 2010, \apj, 721, 98

\bibitem[\protect\citeauthoryear{{Kaviraj}, {} \& {et al.}}{{Kaviraj}
  et~al.}{2012}]{Kaviraj2012}
{Kaviraj} S.,  {}   {et al.} 2012, arXiv:1206.2360

\bibitem[\protect\citeauthoryear{{Kaviraj}, {Tan}, {Ellis} \& {Silk}}{{Kaviraj}
  et~al.}{2011}]{Kaviraj2011}
{Kaviraj} S.,  {Tan} K.-M.,  {Ellis} R.~S.,    {Silk} J.,  2011, \mnras, 411,
  2148

\bibitem[\protect\citeauthoryear{{Kere{\v s}}, {Katz}, {Fardal}, {Dav{\'e}} \&
  {Weinberg}}{{Kere{\v s}} et~al.}{2009}]{Keres2009}
{Kere{\v s}} D.,  {Katz} N.,  {Fardal} M.,  {Dav{\'e}} R.,    {Weinberg} D.~H.,
   2009, \mnras, 395, 160

\bibitem[\protect\citeauthoryear{{Kocevski}, {} \& {et al.}}{{Kocevski}
  et~al.}{2012}]{Kocevski2012}
{Kocevski} D.~D.,  {}   {et al.} 2012, \apj, 744, 148

\bibitem[\protect\citeauthoryear{{Komatsu}, {} \& {et al.}}{{Komatsu}
  et~al.}{2011}]{Komatsu2011}
{Komatsu} E.,  {}   {et al.} 2011, \apjs, 192, 18

\bibitem[\protect\citeauthoryear{{Law}, {Steidel}, {Erb}, {Larkin}, {Pettini},
  {Shapley} \& {Wright}}{{Law} et~al.}{2009}]{Law2009}
{Law} D.~R.,  {Steidel} C.~C.,  {Erb} D.~K.,  {Larkin} J.~E.,  {Pettini} M.,
  {Shapley} A.~E.,    {Wright} S.~A.,  2009, \apj, 697, 2057

\bibitem[\protect\citeauthoryear{{Law}, {Steidel}, {Shapley}, {Nagy}, {Reddy}
  \& {Erb}}{{Law} et~al.}{2012a}]{Law2012}
{Law} D.~R.,  {Steidel} C.~C.,  {Shapley} A.~E.,  {Nagy} S.~R.,  {Reddy} N.~A.,
     {Erb} D.~K.,  2012a, \apj, 745, 85

\bibitem[\protect\citeauthoryear{{Law}, {Steidel}, {Shapley}, {Nagy}, {Reddy}
  \& {Erb}}{{Law} et~al.}{2012b}]{Law2012b}
{Law} D.~R.,  {Steidel} C.~C.,  {Shapley} A.~E.,  {Nagy} S.~R.,  {Reddy} N.~A.,
     {Erb} D.~K.,  2012b, ArXiv e-prints

\bibitem[\protect\citeauthoryear{{Lintott}, {} \& {et al.}}{{Lintott}
  et~al.}{2008}]{Lintott2008}
{Lintott} C.~J.,  {}   {et al.} 2008, \mnras, 389, 1179

\bibitem[\protect\citeauthoryear{{Lisker}}{{Lisker}}{2008}]{Lisker2008}
{Lisker} T.,  2008, ApJS, 179, 319

\bibitem[\protect\citeauthoryear{{Lotz}, {} \& {et al.}}{{Lotz}
  et~al.}{2008}]{Lotz2008}
{Lotz} J.~M.,  {}   {et al.} 2008, ApJ, 672, 177

\bibitem[\protect\citeauthoryear{{Lotz}, {Madau}, {Giavalisco}, {Primack} \&
  {Ferguson}}{{Lotz} et~al.}{2006}]{Lotz2006}
{Lotz} J.~M.,  {Madau} P.,  {Giavalisco} M.,  {Primack} J.,    {Ferguson}
  H.~C.,  2006, ApJ, 636, 592

\bibitem[\protect\citeauthoryear{{Lotz}, {Primack} \& {Madau}}{{Lotz}
  et~al.}{2004}]{Lotz2004}
{Lotz} J.~M.,  {Primack} J.,    {Madau} P.,  2004, AJ, 128, 163

\bibitem[\protect\citeauthoryear{{Madau}, {Pozzetti} \& {Dickinson}}{{Madau}
  et~al.}{1998}]{Madau1998}
{Madau} P.,  {Pozzetti} L.,    {Dickinson} M.,  1998, \apj, 498, 106

\bibitem[\protect\citeauthoryear{{Mancini}, {} \& {et al.}}{{Mancini}
  et~al.}{2011}]{Mancini2011}
{Mancini} C.,  {}   {et al.} 2011, \apj, 743, 86

\bibitem[\protect\citeauthoryear{{Mihos} \& {Hernquist}}{{Mihos} \&
  {Hernquist}}{1996}]{Mihos1996}
{Mihos} J.~C.,  {Hernquist} L., , 1996, {Gasdynamics and Starbursts in Major
  Mergers}

\bibitem[\protect\citeauthoryear{{Newman}, {Ellis}, {Bundy} \& {Treu}}{{Newman}
  et~al.}{2012}]{Newman2012}
{Newman} A.~B.,  {Ellis} R.~S.,  {Bundy} K.,    {Treu} T.,  2012, \apj, 746,
  162

\bibitem[\protect\citeauthoryear{{Oke} \& {Gunn}}{{Oke} \&
  {Gunn}}{1983}]{Oke1983}
{Oke} J.~B.,  {Gunn} J.~E.,  1983, \apj, 266, 713

\bibitem[\protect\citeauthoryear{{Pannella}, {} \& {et al.}}{{Pannella}
  et~al.}{2009}]{Pannella2009}
{Pannella} M.,  {}   {et al.} 2009, \apjl, 698, L116

\bibitem[\protect\citeauthoryear{{Popesso}, {} \& {et al.}}{{Popesso}
  et~al.}{2009}]{Popesso2009}
{Popesso} P.,  {}   {et al.} 2009, \aap, 494, 443

\bibitem[\protect\citeauthoryear{{Reddy}, {Erb}, {Steidel}, {Shapley},
  {Adelberger} \& {Pettini}}{{Reddy} et~al.}{2005}]{Reddy2005}
{Reddy} N.~A.,  {Erb} D.~K.,  {Steidel} C.~C.,  {Shapley} A.~E.,  {Adelberger}
  K.~L.,    {Pettini} M.,  2005, \apj, 633, 748

\bibitem[\protect\citeauthoryear{{Reddy}, {Pettini}, {Steidel}, {Shapley},
  {Erb} \& {Law}}{{Reddy} et~al.}{2012}]{Reddy2012}
{Reddy} N.~A.,  {Pettini} M.,  {Steidel} C.~C.,  {Shapley} A.~E.,  {Erb} D.~K.,
     {Law} D.~R.,  2012, arXiv:1205.0555

\bibitem[\protect\citeauthoryear{{Reddy} \& {Steidel}}{{Reddy} \&
  {Steidel}}{2004}]{Reddy2004}
{Reddy} N.~A.,  {Steidel} C.~C.,  2004, \apjl, 603, L13

\bibitem[\protect\citeauthoryear{{Rodighiero}, {} \& {et al.}}{{Rodighiero}
  et~al.}{2011}]{Rodighiero2011}
{Rodighiero} G.,  {}   {et al.} 2011, \apjl, 739, L40

\bibitem[\protect\citeauthoryear{{Santini}, {} \& {et al.}}{{Santini}
  et~al.}{2009}]{Santini2009}
{Santini} P.,  {}   {et al.} 2009, \aap, 504, 751

\bibitem[\protect\citeauthoryear{{Shapiro}, {} \& {et al.}}{{Shapiro}
  et~al.}{2008}]{Shapiro2008}
{Shapiro} K.~L.,  {}   {et al.} 2008, \apj, 682, 231

\bibitem[\protect\citeauthoryear{{Shapley}, {Steidel}, {Erb}, {Reddy},
  {Adelberger}, {Pettini}, {Barmby} \& {Huang}}{{Shapley}
  et~al.}{2005}]{Shapley2005}
{Shapley} A.~E.,  {Steidel} C.~C.,  {Erb} D.~K.,  {Reddy} N.~A.,  {Adelberger}
  K.~L.,  {Pettini} M.,  {Barmby} P.,    {Huang} J.,  2005, \apj, 626, 698

\bibitem[\protect\citeauthoryear{{Taylor-Mager}, {Conselice}, {Windhorst} \&
  {Jansen}}{{Taylor-Mager} et~al.}{2007}]{Taylor-Mager2007}
{Taylor-Mager} V.~A.,  {Conselice} C.~J.,  {Windhorst} R.~A.,    {Jansen}
  R.~A.,  2007, \apj, 659, 162

\bibitem[\protect\citeauthoryear{{Tresse}, {} \& {et al.}}{{Tresse}
  et~al.}{2007}]{Tresse2007}
{Tresse} L.,  {}   {et al.} 2007, \aap, 472, 403

\bibitem[\protect\citeauthoryear{{van Dokkum}, {} \& {et al.}}{{van Dokkum}
  et~al.}{2011}]{VD2011}
{van Dokkum} P.~G.,  {}   {et al.} 2011, \apjl, 743, L15

\bibitem[\protect\citeauthoryear{{Windhorst}, {} \& {et al.}}{{Windhorst}
  et~al.}{2002}]{Windhorst2002}
{Windhorst} R.~A.,  {}   {et al.} 2002, \apjs, 143, 113

\bibitem[\protect\citeauthoryear{{Windhorst}, {} \& {et al.}}{{Windhorst}
  et~al.}{2011}]{Windhorst2011}
{Windhorst} R.~A.,  {}   {et al.} 2011, \apjs, 193, 27

\bibitem[\protect\citeauthoryear{{Wuyts}, {} \& {et al.}}{{Wuyts}
  et~al.}{2011}]{Wuyts2011}
{Wuyts} S.,  {}   {et al.} 2011, \apj, 738, 106

\end{thebibliography}


\end{document}